\documentclass[twocolumn,traditabstract]{aa}

\usepackage{graphicx,amsmath}
\usepackage{epsf}
\usepackage{txfonts}
\usepackage[hyphens]{url}
\usepackage{longtable}
\usepackage{lscape}
\usepackage[hyperindex,breaklinks=true, colorlinks, citecolor=blue]{hyperref}  
\usepackage{breakurl}
\usepackage{comment}
\usepackage{natbib}
\usepackage{upgreek}
\usepackage{float}
\usepackage[switch,pagewise]{lineno}
\usepackage{multirow}

\newcommand{\hpatrick}[1]{}
\newcommand{\commentproof}[1]{{\bf \color{green}#1}}
\renewcommand{\commentproof}[1]{} 

\def\setsymbol#1#2{\expandafter\def\csname #1\endcsname{#2}}
\def\getsymbol#1{\csname #1\endcsname}

\def\Planck{\textit{Planck}}

\def\all2013resultspapers{\nocite{planck2013-p01, planck2013-p02, planck2013-p02a, planck2013-p02d, planck2013-p02b, planck2013-p03, planck2013-p03c, planck2013-p03f, planck2013-p03d, planck2013-p03e, planck2013-p01a, planck2013-p06, planck2013-p03a, planck2013-pip88, planck2013-p08, planck2013-p11, planck2013-p12, planck2013-p13, planck2013-p14, planck2013-p15, planck2013-p05b, planck2013-p17, planck2013-p09, planck2013-p09a, planck2013-p20, planck2013-p19, planck2013-pipaberration, planck2013-p05, planck2013-p05a, planck2013-pip56, planck2013-p06b, planck2013-p01a}}

\newbox\tablebox    \newdimen\tablewidth
\def\leaderfil{\leaders\hbox to 5pt{\hss.\hss}\hfil}
\def\tablenote#1 #2\par{\begingroup \parindent=0.8em
    \abovedisplayshortskip=0pt\belowdisplayshortskip=0pt
    \noindent
    $$\hss\vbox{\hsize\tablewidth \hangindent=\parindent \hangafter=1 \noindent
    \hbox to \parindent{$^#1$\hss}\strut#2\strut\par}\hss$$
    \endgroup}

\def\L2{\ifmmode L_2\else $L_2$\fi}

\def\DeltaT{\ifmmode \Delta T\else $\Delta T$\fi}
\def\deltat{\ifmmode \Delta t\else $\Delta t$\fi}
\def\fknee{\ifmmode f_{\rm knee}\else $f_{\rm knee}$\fi}
\def\Fmax{\ifmmode F_{\rm max}\else $F_{\rm max}$\fi}
\def\solar{\ifmmode{\rm M}_{\mathord\odot}\else${\rm M}_{\mathord\odot}$\fi}
\def\Msolar{\ifmmode{\rm M}_{\mathord\odot}\else${\rm M}_{\mathord\odot}$\fi}
\def\Lsolar{\ifmmode{\rm L}_{\mathord\odot}\else${\rm L}_{\mathord\odot}$\fi}
\def\inv{\ifmmode^{-1}\else$^{-1}$\fi}
\def\mo{\ifmmode^{-1}\else$^{-1}$\fi}
\def\sup#1{\ifmmode ^{\rm #1}\else $^{\rm #1}$\fi}
\def\expo#1{\ifmmode \times 10^{#1}\else $\times 10^{#1}$\fi}
\def\,{\thinspace}
\def\lsim{\mathrel{\raise .4ex\hbox{\rlap{$<$}\lower 1.2ex\hbox{$\sim$}}}}
\def\gsim{\mathrel{\raise .4ex\hbox{\rlap{$>$}\lower 1.2ex\hbox{$\sim$}}}}

\def\simprop{\mathrel{\raise .4ex\hbox{\rlap{$\propto$}\lower 1.2ex\hbox{$\sim$}}}}
\def\deg{\ifmmode^\circ\else$^\circ$\fi}
\def\pdeg{\ifmmode $\setbox0=\hbox{$^{\circ}$}\rlap{\hskip.11\wd0 .}$^{\circ}
          \else \setbox0=\hbox{$^{\circ}$}\rlap{\hskip.11\wd0 .}$^{\circ}$\fi}
\def\arcs{\ifmmode {^{\scriptstyle\prime\prime}}
          \else $^{\scriptstyle\prime\prime}$\fi}
\def\arcm{\ifmmode {^{\scriptstyle\prime}}
          \else $^{\scriptstyle\prime}$\fi}
\newdimen\sa  \newdimen\sb
\def\parcs{\sa=.07em \sb=.03em
     \ifmmode \hbox{\rlap{.}}^{\scriptstyle\prime\kern -\sb\prime}\hbox{\kern -\sa}
     \else \rlap{.}$^{\scriptstyle\prime\kern -\sb\prime}$\kern -\sa\fi}
\def\parcm{\sa=.08em \sb=.03em
     \ifmmode \hbox{\rlap{.}\kern\sa}^{\scriptstyle\prime}\hbox{\kern-\sb}
     \else \rlap{.}\kern\sa$^{\scriptstyle\prime}$\kern-\sb\fi}
\def\ra[#1 #2 #3.#4]{#1\sup{h}#2\sup{m}#3\sup{s}\llap.#4}
\def\dec[#1 #2 #3.#4]{#1\deg#2\arcm#3\arcs\llap.#4}
\def\deco[#1 #2 #3]{#1\deg#2\arcm#3\arcs}
\def\rra[#1 #2]{#1\sup{h}#2\sup{m}}

\def\dots{\relax\ifmmode \ldots\else $\ldots$\fi}
\def\WHzsr{\ifmmode $W\,Hz\mo\,sr\mo$\else W\,Hz\mo\,sr\mo\fi}
\def\mHz{\ifmmode $\,mHz$\else \,mHz\fi}
\def\GHz{\ifmmode $\,GHz$\else \,GHz\fi}
\def\mKs{\ifmmode $\,mK\,s$^{1/2}\else \,mK\,s$^{1/2}$\fi}
\def\muKs{\ifmmode \,\mu$K\,s$^{1/2}\else \,$\mu$K\,s$^{1/2}$\fi}
\def\muKRJs{\ifmmode \,\mu$K$_{\rm RJ}$\,s$^{1/2}\else \,$\mu$K$_{\rm RJ}$\,s$^{1/2}$\fi}
\def\muKHz{\ifmmode \,\mu$K\,Hz$^{-1/2}\else \,$\mu$K\,Hz$^{-1/2}$\fi}
\def\MJysr{\ifmmode \,$MJy\,sr\mo$\else \,MJy\,sr\mo\fi}
\def\MJysrmK{\ifmmode \,$MJy\,sr\mo$\,mK$_{\rm CMB}\mo\else \,MJy\,sr\mo\,mK$_{\rm CMB}\mo$\fi}
\def\microns{\ifmmode \,\mu$m$\else \,$\mu$m\fi}
\def\micron{\microns}
\def\muK{\ifmmode \,\mu$K$\else \,$\mu$\hbox{K}\fi}
\def\microK{\ifmmode \,\mu$K$\else \,$\mu$\hbox{K}\fi}
\def\muW{\ifmmode \,\mu$W$\else \,$\mu$\hbox{W}\fi}
\def\kms{\ifmmode $\,km\,s$^{-1}\else \,km\,s$^{-1}$\fi}
\def\kmsMpc{\ifmmode $\,\kms\,Mpc\mo$\else \,\kms\,Mpc\mo\fi}
\providecommand{\sorthelp}[1]{}

\include{polar_definitions}

\def\Herschel{\textit{Herschel}}
\newcommand{\nh}{$N_{\textsc{H}}$}

\providecommand{\sorthelp}[1]{}

\newcommand{\bperp}{$\langle\hat{\vec{B}}_{\perp}\rangle$}
\newcommand{\wc}{{\mkern 2mu\cdot\mkern 2mu}}

\begin{document}

\title{What are we learning from the relative orientation between density structures and the magnetic field in molecular clouds?}
\titlerunning{Relative orientation between density structures and magnetic field}
\author{J.~D.~Soler$^{1,2}$ \and P.~Hennebelle$^{2}$\\}    
\institute{
1. Max-Planck-Institute for Astronomy, K\"{o}nigstuhl 17, 69117, Heidelberg, Germany. \email{soler@mpia.de}\\
2. Laboratoire AIM, Paris-Saclay, CEA/IRFU/SAp - CNRS - Universit\'{e} Paris Diderot, 91191, Gif-sur-Yvette Cedex, France}
\authorrunning{Soler,\,J.D. \& Hennebelle\,P.}

\date{Received 26 April 2017 / Accepted nn XX 201X}

\abstract{We investigate the conditions of ideal magnetohydrodynamic (MHD) turbulence responsible for the relative orientation between density structures, characterized by their gradient, $\vec{\nabla}\rho$, and the magnetic field, $\vec{B}$, in molecular clouds (MCs).
For that purpose, we construct an expression for the time evolution of the angle, $\phi$, between $\vec{\nabla}\rho$ and $\vec{B}$ based on the transport equations of MHD turbulence.
Using this expression, we find that the configuration where $\vec{\nabla}\rho$ and $\vec{B}$ are mostly parallel, $\cos\phi=1$, and where $\vec{\nabla}\rho$ and $\vec{B}$ are mostly perpendicular, $\cos\phi=0$, constitute attractors, that is, the system tends to evolve towards either of these configurations and they are more represented than others.
This fact would explain the predominant alignment or anti-alignment between column density, \nh, structures and the projected magnetic field orientation, \bperp, reported in observations.
Additionally, we find that departures from the $\cos\phi=0$ configurations are related to convergent flows, quantified by the divergence of the velocity field, $\vec{\nabla}\wc\vec{v}$, in the presence of a relatively strong magnetic field.
This would explain the observed change in relative orientation between \nh-structures and \bperp\ towards MCs, from mostly parallel at low \nh\ to mostly perpendicular at the highest \nh, as the result of the gravitational collapse and/or convergence of flows.
Finally, we show that the density threshold that marks the observed change in relative orientation towards MCs, from \nh\ and \bperp\ being mostly parallel at low \nh\ to mostly perpendicular at the highest \nh, is related to the magnetic field strength and constitutes a crucial piece of information for determining the role of the magnetic field in the dynamics of MCs.}
\keywords{ISM: general, structure, magnetic fields, clouds -- Magnetohydrodynamics (MHD) -- Turbulence}

\maketitle

\section{Introduction}\label{section:introduction}

The advent of novel observations of polarization from dust, in emission at submillimeter wavelengths and in absorption from background stars in the visible and near-infrared, provides an unprecedented amount of information on the magnetic field orientation integrated along the line of sight and projected on the plane of the sky, \bperp, at molecular cloud (MC) scales \citep{crutcher2012,planck2014-a01}.
However, integrating this information to our understanding of the dynamical processes that produce density structures in the interstellar medium (ISM), from MCs to filaments and eventually stars, remains challenging \citep{bergin2007,hennebelle2012,klessen2016}.

One simple approach to obtain insight into the role of the magnetic field is the study of the relative orientation between the observed column density, \nh, structures and \bperp. 
Multiple studies of \bperp\ inferred from starlight polarization, e.g., \cite{palmeirim2013, li2013, sugitani2011} and more recently \cite{kusune2016, santos2016a, soler2016, hoq2017}, use qualitative descriptions of the relative orientation between \bperp\ and the column density structures to argue the importance of the magnetic field in structuring the observed regions.

In emission from the diffuse ISM, quantitative analysis of the polarization observations at 353\,GHz by ESA's \Planck\ satellite show that over most of the sky, the majority of the elongated column density structures traced by dust thermal emission are predominantly aligned with the magnetic field measured on the structures \citep{planck2014-XXXII}.
This statistical trend, which becomes less striking for increasing column density, is similar to that found between low column density ($N_{\rm H\text{\sc i}}\approx5\times10^{18}$\,cm$^{-2}$) fibres traced by H{\sc i} emission and \bperp\ \citep{clark2014,kalberla2016}.

In emission from the denser ISM, quantitative analysis of the \Planck\ polarization observations at 353\,GHz towards ten nearby ($d < 450$\,pc) MCs shows that the relative orientation between the column density structures and \bperp\ progressively changes with increasing column density from mostly parallel at $\log(N_{\rm H}/\text{cm}^{-2})\lsim21.7$ to mostly perpendicular at $\log(N_{\rm H}/\text{cm}^{-2})\gsim21.7$ \citep{planck2015-XXXV}.
Subsequent studies of the relative orientation between \nh\ structures and \bperp\ have identified similar trends using \nh\ structures derived from \Herschel\ observations at 20\arcsec\ resolution and \bperp\ inferred from BLASTPol polarization observation at 250, 350, and 500\micron\ towards the Vela C molecular complex \citep{soler2017} as well as using \nh\ structures derived from \Herschel\ observations together with \Planck\ 353\,GHz polarization observations towards the high-latitude cloud L1642 \citep{malinen2016}.

From the theoretical point of view, the magnetic field, whose observed energy density is in rough equipartition with other energy densities in the local ISM \citep{heiles2005}, imposes an asymmetry for the formation of condensations from the diffuse ISM.
Condensation modes are unaffected when they propagate parallel to the mean direction of the field, $\vec{B}_{0}$, but inhibited by the magnetic pressure when they propagate normal to $\vec{B}_{0}$ \citep{field1965}.
Consequently, condensations driven by thermal instability arise along $\vec{B}_{0}$ and MCs can be formed by shock waves only if the perturbations propagate almost parallel to the mean magnetic field \citep{hennebelle2000,hartmann2001,inoue2007,kortgen2015}.

This theoretical framework suggests that the observed regimes in relative orientation between \nh\ structures and \bperp\ indicate something fundamental about the gathering of gas out of the diffuse ISM, which results in the formation of MCs and their subsequent evolution to form denser structures, such as filaments and cores.
In this work, we explore the relation between the observed relative orientation between \bperp\ and the \nh\ structures and the conditions imposed by the transport equations of magnetohydrodynamic (MHD) turbulence.
For that purpose, we construct an expression for the evolution of the relative orientation between $\vec{\nabla}\rho$ and $\vec{B}$ and evaluate the physical processes that are potentially responsible for the observed trends.

This paper is organized as follows.
In Section~\ref{sec:equations} we introduce the transport equations of MHD turbulence and derive an expression for the relative orientation between $\vec{\nabla}\rho$ and \vec{B}.
In Section~\ref{sec:interpretation} we discuss the implications of the derived relation and characterize it using, first, a set of simple cases, and, second, the simulations of MHD turbulence in MCs that were used to characterize the analysis method presented in \cite{planck2015-XXXV}.
Section~\ref{sec:discussion} discusses the implications of the studied relation between $\vec{\nabla}\rho$ and $\vec{B}$.
Finally, Section~\ref{sec:conclusions} gives our conclusions and anticipates future work.
We reserve some additional analyses to Appendix~\ref{app:BandV}, where we present an expression for time evolution of the relative orientation between the magnetic, $\vec{B}$, and the velocity, $\vec{v}$, fields.

\section{Relative orientation between \vec{B} and iso-density contours}\label{sec:equations}

\subsection{Transport equations of magnetohydrodynamic turbulence}\label{sec:MHDeq}

In the Lagrangian specification, which corresponds to looking at the fluid motion following an individual parcel as it moves through space and time, the continuity equation, which guarantees the conservation of mass, is
\begin{equation}\label{eq:continuity}
\frac{d\log\rho}{d t}=-\partial_{j}v_{j},
\end{equation}
where $\rho$ and $v_{j}$ respectively being the density and velocity, and with the index $i$ running between $0$ and $2$ representing the axes of a 3D Cartesian reference frame.
Here, and in the rest of the paper, summation over repeated indexes is implied, following the Einstein summation convention

The magnetic field, $\vec{B}$, is described by the Faraday equation
\begin{equation}\label{eq:Faraday}
\partial_{t}\vec{B}=\vec{\nabla}\times(\vec{v}\times\vec{B}),
\end{equation}
which can be reduced, under the assumption of low magnetic diffusivity, to the magnetic induction equation
\begin{equation}\label{eq:Faraday1}
\frac{dB_{i}}{d t} = B_{j}(\partial_{j}v_{i}) - B_{i}(\partial_{j}v_{j}).
\end{equation}

\subsection{Time evolution of the relative orientation between $\vec{\nabla}\rho$ and $\vec{B}$}\label{sec:BvsV} 

We construct an expression for behaviour of $\cos\phi$, the angle between $\vec{\nabla}\rho$ and $\vec{B}$, by combining the equations introduced in Section~\ref{sec:MHDeq} as follows.
By the definition of the scalar product of vectors, the cosine of the angle between $\vec{\nabla}\rho$ and $\vec{B}$ corresponds to
\begin{equation}\label{eq:relorientation0}
\begin{split}
\cos\phi &= \frac{\vec{\nabla}\rho\wc\vec{B}}{|\vec{\nabla}\rho|\,|\vec{B}|} \\
 	     &= \frac{R_{i}B_{i}}{(R_{j}R_{j})^{1/2}(B_{k}B_{k})^{1/2}},    
\end{split}	     
\end{equation}
where we introduce the convention $R_{i}\equiv\partial_{i}\log\rho$.
Note that the distribution of relative orientations between two sets of uniformly-distributed random vectors in 3D is flat in the cosine of their separation angle, thus all the discussions of relative orientations in 3D are given in terms of $\cos\phi$ \citep[see appendix~C of][]{planck2015-XXXV}.

We apply the time derivative to the square of Eq.~\ref{eq:relorientation0} and, assuming that the spatial and time derivatives can be commuted, we obtain
\begin{equation}\label{eq:costheta0}
\begin{split}
\frac{d(\cos^{2}\phi)}{dt} &= \frac{2}{(R_{k}R_{k})^{2}(B_{k}B_{k})^{2}}\,\bigg[\frac{d(R_{i}B_{i})}{d t}(R_{i}B_{i})(R_{m}R_{m})(B_{n}B_{n}) \\
&-\frac{d(B_{i})}{d t}B_{i}(R_{j}B_{j})^{2}(R_{m}R_{m}) -\frac{d(R_{i})}{d t}R_{i}(B_{n}B_{n})(R_{j}B_{j})^{2}\bigg],
\end{split}	   
\end{equation}
where we use the definitions 
\begin{equation}\label{eq:defri}
r_{i} \equiv \frac{R_{i}}{(R_{k}R_{k})^{1/2}},
\end{equation}
\begin{equation}\label{eq:defbi}
b_{i} \equiv \frac{B_{i}}{(B_{k}B_{k})^{1/2}},
\end{equation}
which in combination with Eq.~\ref{eq:relorientation0} lead to
\begin{equation}\label{eq:costheta1}
\begin{split}
\frac{d(\cos\phi)}{dt} =& \frac{1}{(R_{k}R_{k})^{1/2}(B_{k}B_{k})^{1/2}}\,\frac{d(R_{i}B_{i})}{d t} \\
&-\Bigg[\frac{b_{i}}{(B_{k}B_{k})^{1/2}}\,\frac{d(B_{i})}{d t}+\frac{r_{i}}{(R_{k}R_{k})^{1/2}}\,\frac{d(R_{i})}{d t}\Bigg]\cos\phi, \\
=& [r_{i}-b_{i}\cos\phi]\frac{1}{(B_{k}B_{k})^{1/2}}\,\frac{d(B_{i})}{d t} \\
&+ [b_{i}-r_{i}\cos\phi]\frac{1}{(R_{k}R_{k})^{1/2}}\,\frac{d(R_{i})}{d t}.
\end{split}	   
\end{equation}

In the particular case where $\cos\phi=\pm1$, which corresponds to $r_{i}=\pm b_{i}$, Eq.~\ref{eq:costheta1} becomes $d(\cos\phi)/dt = 0$, thus implying that this configuration is an attractor, that is, a configuration towards which the system tends to evolve for a wide variety of starting conditions.
This attractor is generic in the sense that it is purely geometrical and does not depend on the details of the physics, as long as the time derivatives in Eq.~\ref{eq:costheta1} do not become infinite when $\cos\phi=\pm1$. 

From Eq.~\ref{eq:continuity} we get
\begin{equation}\label{eq:continuity2}
\begin{split}
\partial_{i}\left(\frac{d(\log\rho)}{d t}\right) &= -\partial_{i}[\partial_{j}(\log\rho)v_{j}+\log\rho(\partial_{j}v_{j})],
\end{split}	
\end{equation}
which is equivalent to
\begin{equation}\label{eq:continuity3}
\begin{split}
\frac{d(R_{i})}{d t} &=  -\partial_{i}(\partial_{j}v_{j}) - (\partial_{i}v_{j})R_{j}. 
\end{split}	
\end{equation}

Finally, we introduce Eq.~\ref{eq:Faraday1} and Eq.~\ref{eq:continuity3} into Eq.~\ref{eq:costheta1} to obtain 
\begin{equation}\label{eq:costheta3}
\frac{d(\cos\phi)}{dt} = \frac{\partial_{i}(\partial_{j}v_{j})}{(R_{k}R_{k})^{1/2}}[-b_{i}+r_{i}\cos\phi]+\partial_{i}v_{j}[r_{i}r_{j}-b_{i}b_{j}]\cos\phi,
\end{equation}
which corresponds to the time evolution of the cosine of the angle between $\vec{\nabla}\rho$ and $\vec{B}$.
Note that this expression is entirely based on the transport equations of ideal MHD turbulence.
It is composed by the strain tensor, $\partial_{i}v_{j}$ \citep{landau1959}, and the symmetric tensors $r_{i}r_{j}$ and $b_{i}b_{j}$, which represent the correlation of the density gradient orientation and the correlation between the components of the magnetic field orientation, respectively.

For the sake of simplicity, in the rest of this document we rewrite Eq.~\ref{eq:costheta3} as
\begin{equation}\label{eq:costhetaCoeff}
\frac{d(\cos\phi)}{dt} = C+ [A_{1}+A_{23}]\cos\phi,
\end{equation}
where we use the definitions
\begin{equation}\label{eq:A1}
A_{1} \equiv \frac{\partial_{i}(\partial_{j}v_{j})}{(R_{k}R_{k})^{1/2}}r_{i},
\end{equation}
\begin{equation}\label{eq:C}
C \equiv -\frac{\partial_{i}(\partial_{j}v_{j})}{(R_{k}R_{k})^{1/2}}b_{i},
\end{equation}
and 
\begin{equation}\label{eq:A23}
\begin{split}
A_{23} &\equiv \partial_{i}v_{j}[r_{i}r_{j}-b_{i}b_{j}],
\end{split}	
\end{equation}
which, without any loss in generality, can be expressed as
\begin{equation}\label{eq:A23sym}
\begin{split}
A_{23} &\equiv \frac{1}{2}(\partial_{i}v_{j}+\partial_{j}v_{i})[r_{i}r_{j}-b_{i}b_{j}]\,.
\end{split}	
\end{equation}

For the sake of comparison with previous works that studied the relative orientation between $\vec{B}$ and $\vec{v}$ \citep{matthaeus2008,banerjee2009}, we obtained a similar expression for the relative orientation between those two quantities and present it in Appendix~\ref{app:BandV}.

\section{Interpretation}\label{sec:interpretation}

A few points can be readily concluded from Eq.~\ref{eq:costheta3}.
First, in the case where the coefficient $C$ is very small, which is in general the case as it is composed of second-order spatial derivatives whose value is small compared to the other terms, $\cos\phi=0$ constitutes another attractor.
This means that, under the assumptions presented in Sec.~\ref{sec:equations}, we expect either $\cos\phi=\pm1$ or $\cos\phi=0$, that is, $\vec{B}$ tends to be either parallel or perpendicular to the density structures.
Second, the relative orientation between $\vec{\nabla}\rho$ and $\vec{B}$ changes by the effect of the coupling between the motions of the fluid represented by the strain tensor, $\partial_{i}v_{j}$ and the symmetric unitary tensors $r_{i}r_{j}$ and $b_{i}b_{j}$.
To better understand the implications of Eq.~\ref{eq:costheta3}, we present the study a of few simple cases in Sec.~\ref{sec:SimpleExamples} and of a set of simulations of MHD turbulence in Sec.~\ref{sec:MHDExample}.

\subsection{Study of simples cases}\label{sec:SimpleExamples}

In order to develop some intuition into the information encompassed in Eq.~\ref{eq:costheta3}, we study its behaviour in a few simple distributions of matter, velocity, and magnetic field.

\subsubsection{Strong magnetic field}

If we consider a very strong magnetic field, such that $\vec{B}$ is oriented almost exclusively along one particular axis, and a converging flow along the same direction, Eq.~\ref{eq:costheta3} reduces to
\begin{equation}
\frac{d(\cos\phi)}{dt} \simeq \partial_{0}v_{0}(r_{0}r_{0}-1)\cos\phi,
\end{equation}
where we ignore the terms depending on $\partial_{i}(\partial_{j}v_{j})$, which are much smaller than those related to $\partial_{i}v_{j}$.
If the matter is initially distributed mostly parallel to $\vec{B}$, $r_{0}r_{0}\approx0$ and given that $\partial_{0}v_{0} < 0$ , Eq.~\ref{eq:costheta3} implies that $\cos\phi$ increases, that is, the relative orientation changes from its initial configuration.
Alternatively, if the matter is initially distributed mostly perpendicular to $\vec{B}$, $r_{0}r_{0}\approx1$ and Eq.~\ref{eq:costheta3} indicates that there is no change in relative orientation between $\vec{\nabla}\rho$ and $\vec{B}$.

If we consider the same configuration of $\vec{B}$, but this time consider a convergent flow perpendicular to it, Eq.~\ref{eq:costheta3} reduces to
\begin{equation}
\frac{d(\cos\phi)}{dt} \simeq \partial_{1}v_{1}(r_{1}r_{1})\cos\phi.
\end{equation}
In this case, if the matter is initially distributed mostly parallel to $\vec{B}$, Eq.~\ref{eq:costheta3} indicates that there is no change in relative orientation and the matter would continue to be stretched along the magnetic field.
Alternatively, if the matter is initially distributed mostly perpendicular to $\vec{B}$, $r_{1}r_{1}\approx1$ and given that $\partial_{0}v_{0} < 0$ , Eq.~\ref{eq:costheta3} implies that $\cos\phi$ increases, that is, the relative orientation changes from its initial configuration.
Both of these results confirm that the combination of a strong magnetic field and a convergent flow only results in accumulation of matter when the flow is directed along the magnetic field.

\subsubsection{Weak magnetic field}

If we consider a weak magnetic field, matter distributed in a slab perpendicular to the $x$-axis, and a converging flow mostly parallel to the $x$-axis, Eq.~\ref{eq:costheta3} reduces to 
\begin{equation}\label{eq:wf1}
\frac{d(\cos\phi)}{dt} \simeq \partial_{i}v_{j}(r_{0}r_{0}-b_{i}b_{j})\cos\phi.
\end{equation}
This expression shows that in the case of single compressive flow, the change in relative orientation is the result of two competing terms. The first one, $r_{0}r_{0}$, which indicates the geometrical distribution of $\rho$, and the second one, $b_{i}b_{j}$, which indicates the distribution of the magnetic field orientation and tends to be larger when the field is stronger.
Given that here $\partial_{0}v_{0} <0$, converging flow, and in the case of a weak field $b_{i}b_{i}$ is on average equal to $1/3$, Eq.~\ref{eq:wf1} reduces to
\begin{equation}\label{eq:wf2}
\frac{d(\cos\phi)}{dt} \propto -\cos\phi,
\end{equation}
which has solutions that tend towards $\cos\phi=0$.

Alternatively, if we consider the same $\rho$ and $\vec{B}$ configurations as in the previous example, but this time a converging flow in the $y$-direction, Eq.~\ref{eq:costheta3} reduces to
\begin{equation}\label{eq:wf3}
\frac{d(\cos\phi)}{dt} \simeq -\partial_{1}v_{1}b_{1}b_{1}\cos\phi.
\end{equation}
This expression shows that, in this particular example, the relative orientation between $\vec{\nabla}\rho$ and $\vec{B}$ changes by effect of the $y$-component of the magnetic field. 
Thus, it implies that even if the magnetic field is weak, motions strictly along the field lines tend to create density structures perpendicular to it.

\subsubsection{Analytical hints on the relative orientation}\label{sec:ToyDiagonalized}

The term $A_{23}$, is the result of the contraction of two symmetric tensors and thus can be diagonalized.
If we consider the base were the strain tensor is diagonal, Eq.~\ref{eq:A23sym} can be expressed as 
\begin{equation}\label{eq:A23diag}
A_{23} = \lambda_{i}(r_{i}^2 - b_{i}^2)\,,
\end{equation}
where $\lambda_{i}$ are the eigenvalues of the strain tensor.

For the sake of illustration, we can reduce the problem to 2D and consider two eigenvalues, one negative ($\lambda_{\rm c}$), which is dominant if we consider the case where $\partial_{i}v_{i} <0$, and one positive ($\lambda_{\rm s}$).
In general terms, the fluid parcel is compressed in the direction associated with $\lambda_{\rm c}$ and stretched in the other. 
It can therefore be represented as an ellipsoid whose short axis corresponds to the direction associated to $\lambda_{\rm c}$ and whose major one to the direction associated to $\lambda_{\rm s}$.  
Given that $r_{i}$ is a gradient, it is larger along the short axis, that is, the direction associated to $\lambda_{\rm c}$.

If the magnetic field is weak, $\vec{B}$ tends to be parallel to the major axis of the ellipsoid, because the field is compressed by the converging flow.
That means that $b_{\rm c}$ is small and $A_{23}$ is negative, thus taking the system towards the $\cos\phi=0$ configuration.
If the field is strong, the compression occurs mainly along the field lines, thus making $\vec{B}$ parallel to the main axis of the ellipsoid.
In such case $A_{23} \simeq \lambda_{\rm c}(r_i^2 - b_i^2)$, so the sign depends on the values of $b_i$, which can take the system towards the $\cos\phi=\pm1$ configuration. 

These simple cases correspond to highly idealized flow and magnetic field configurations.
To study configurations where the magnetic field is not infinitively rigid or the flow has not only a single compressive component, we need to consider a realization of a turbulent flow in a molecular cloud, which is accessible through the numerical simulation of MHD turbulence.

\begin{figure}[ht!]
\centerline{\includegraphics[width=0.5\textwidth,angle=0,origin=c]{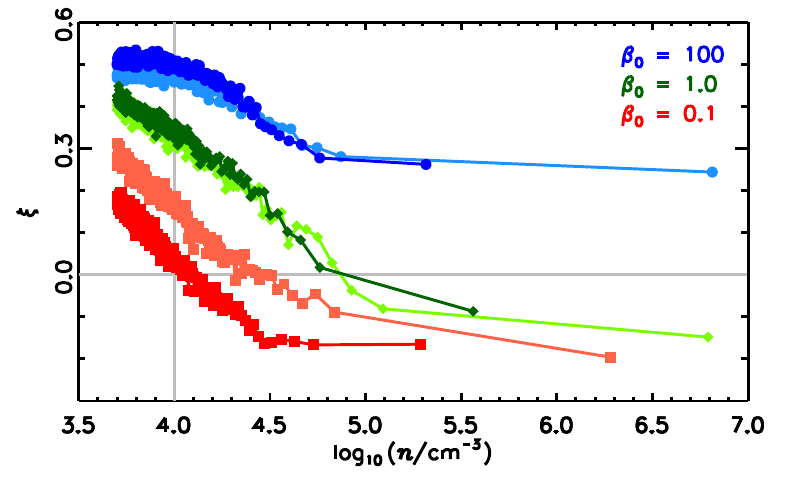}}
\vspace{-0.4cm}
\caption{Relative orientation parameter, $\xi$, as a function of particle density, $n\equiv\rho/\mu$, in the simulations used in \cite{soler2013}.
The values of $\xi$ correspond to the relative orientation between $\vec{\nabla}\rho$ and $\vec{B}$ in $n$-bins with equal number of voxels, all with $n>500$\,cm$^{-3}$.
The values $\xi > 0$ correspond to $\vec{\nabla}\rho$ mostly perpendicular to $\vec{B}$ and $\xi < 0$ correspond to $\vec{\nabla}\rho$ mostly parallel to $\vec{B}$.
The grey horizontal line is $\xi=0$, which corresponds to the case where there is no preferred relative orientation between $\vec{\nabla}\rho$ and $\vec{B}$.
The colours and the symbols represent the initial magnetization values quantified by $\beta_{0}$.
The darker colours represent the early snapshots in the simulation and the lighter colours represent the later snapshots.
The grey vertical line, drawn for reference, corresponds to $n=10^4$\,cm$^{-3}$.
}
\label{fig:zeta-lognh}
\end{figure}

\begin{figure}[ht!]
\centerline{\includegraphics[width=0.5\textwidth,angle=0,origin=c]{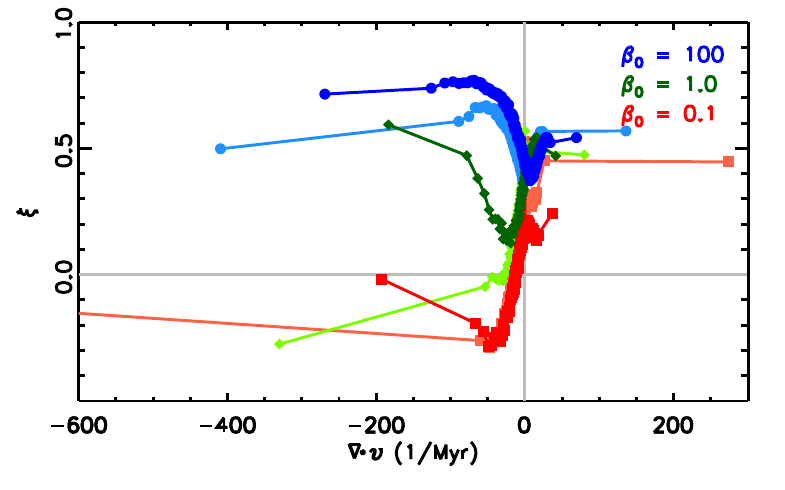}}
\vspace{-0.4cm}
\caption{
Relative orientation parameter, $\xi$, as a function the velocity divergence, $\vec{\nabla}\wc\vec{v}\equiv\partial_{i}v_{i}$, in the simulations introduced in \cite{soler2013}.
The values of $\xi$ correspond to the relative orientation between $\vec{\nabla}\rho$ and $\vec{B}$ in $\vec{\nabla}\wc\vec{v}$-bins with equal number of voxels, all with $n>500$\,cm$^{-3}$.
The values $\xi > 0$ correspond to $\vec{\nabla}\rho$ mostly perpendicular to $\vec{B}$ and $\xi < 0$ correspond to $\vec{\nabla}\rho$ mostly parallel to $\vec{B}$.
The grey horizontal line is $\xi=0$, which corresponds to the case where there is no preferred relative orientation between $\vec{\nabla}\rho$ and $\vec{B}$.
The colours and the symbols represent the initial magnetization values quantified by $\beta_{0}$.
The darker colours represent the early snapshots in the simulation and the lighter colours represent the later snapshots.
The grey vertical line, drawn for reference, corresponds to $\vec{\nabla}\wc\vec{v}=0$\,Myr$^{-1}$.
}
\label{fig:zeta-divV}
\end{figure}

\subsection{Test on a simulation of magnetohydrodynamic turbulence}\label{sec:MHDExample}

We consider the simulations of MHD turbulence introduced in \cite{dib2010} and used in \cite{soler2013}.
These simulations correspond to a 4-parsec-side periodic box with mean number density $n=536$\,cm$^{-3}$, and include the effect of self-gravity, magnetic field, and decaying turbulence.
The medium inside of the box is isothermal ($T=11.4$\,K) and has an initial sonic Mach number $\mathcal{M}_{\rm S}=10$.
These simulations were computed in an adaptive-mesh-refinement (AMR) grid with maximum resolution of $2^{-9}$\,pc and we analyze them in a regular grid with $2^{-7}$\,pc resolution.
For the sake of simplicity, we consider only two snapshots taken at $1/3$ and $2/3$ of the flow crossing time.

This set of simulations includes realizations with three initial degrees of magnetization, quantified in terms of the ratio of the thermal to magnetic pressure, $\beta$: quasi-hydrodynamic, $\beta_{0}=100$; equipartition, $\beta_{0}=1.0$; and strong magnetic field, $\beta_{0}=0.1$.
\cite{soler2013} reported that in 3D, the change in the relative orientation between the magnetic field $\vec{B}$ and the iso-density contours, inferred from $\vec{\nabla}\rho$, is related to the initial degree of magnetization.
In the realizations with $\beta_{0}=0.1$ and $\beta_{0}=1.0$, which correspond to sub-Alfv\'{e}nic or close to equipartition turbulence, $\cos\phi$ changes from being mostly zero at low densities to being mostly plus or minus one at the highest densities.
In the realization with $\beta_{0}=100$, super-Alfv\'{e}nic turbulence, $\cos\phi$ is mostly zero at all densities.
Both of those results were expressed in terms of the relative orientation parameter, $\xi$, which corresponds to the difference between the number of voxels where $\cos\phi\approx0$ minus the number of voxels where $\cos\phi\approx\pm1$ divided by the total number of voxels where $\cos\phi\approx0$ or $\cos\phi\approx\pm1$, as explicitly described in equation 4 of \cite{planck2015-XXXV}.
Consequently, $\xi$ is positive if $\cos\phi$ is mostly equal to zero, that is, $\vec{\nabla}\rho$ mostly perpendicular to $\vec{B}$, and negative if $|\cos\phi|$ is mostly one, that is, $\vec{\nabla}\rho$ mostly parallel to $\vec{B}$.

In order to illustrate the interpretation of Eq.~\ref{eq:costheta3}, we reproduce the relative orientation between $\vec{B}$ and the iso-$\rho$ contours presented in \cite{soler2013} for the range of densities $n>5\times10^{3}$\,cm$^{-3}$.
We estimated $\vec{\nabla}\rho$ using a Lagrange 5-points interpolation to express each $\rho$ data point in the simulation cube as a point on a polynomial and then differentiate that polynomial.\footnote{{\tt pdiv.pro} routine developed by Chris Beaumont (\url{https://github.com/ChrisBeaumont}).}
The mean values of $\xi$ in different density bins, presented in Fig.~\ref{fig:zeta-lognh}, illustrate the different trends in relative orientation between $\vec{\nabla}\rho$ and $\vec{B}$ for different initial magnetizations.
Given that $n\equiv\rho/\mu$ and $\mu$, the mean particle mass, is constant in the simulations, we choose to report these results in terms of $n$ without any loss of generality.
        
Fig.~\ref{fig:zeta-divV} shows that in the same density range, the negative values of $\xi$ are associated with $\vec{\nabla}\wc\vec{v}  < 0$ in the simulations with $\beta_{0}=0.1$ and $\beta_{0}=1.0$.
However, this is not the case for the $\beta_{0}=100$ simulations, thus showing that $\vec{\nabla}\wc\vec{v}  < 0$ is not the only condition producing the change in the relative orientation between $\vec{\nabla}\rho$ parallel to $\vec{B}$.
Furthermore, Fig.~\ref{fig:zeta-divV} illustrates that in the simulations with $\beta_{0}=0.1$ and $\beta_{0}=1.0$, the transition between $\xi > 0$, or $\cos\phi=0$, and $\xi < 0$, or $\cos\phi=\pm1$, happens across $\vec{\nabla}\wc\vec{v}=0$, but $\xi$ is not strictly negative until the second snapshot, when the values of $\vec{\nabla}\wc\vec{v}$ are more negative.

\begin{figure}[ht!]
\centerline{\includegraphics[width=0.5\textwidth,angle=0,origin=c]{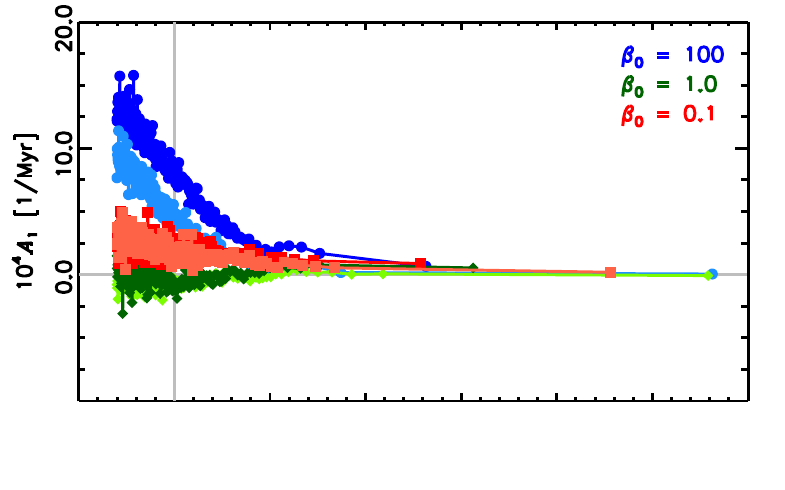}}
\vspace{-1.2cm}
\centerline{\includegraphics[width=0.5\textwidth,angle=0,origin=c]{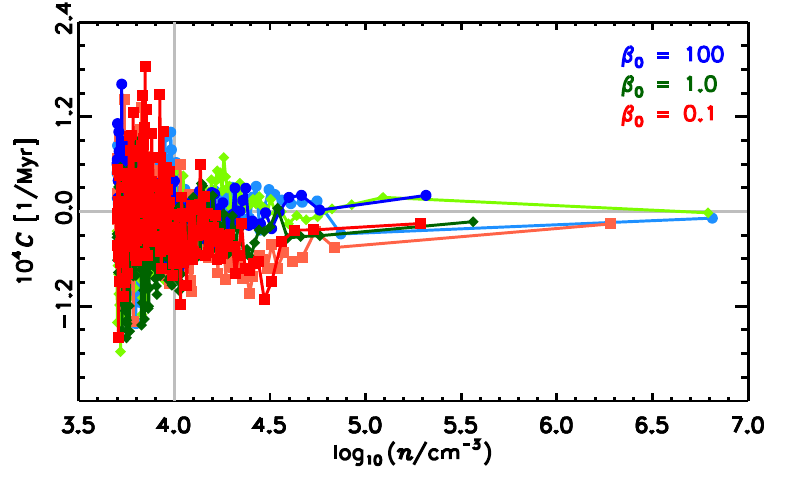}}
\vspace{-0.3cm}
\caption{Mean values of $10^{4}A_{1}$ (top) and $10^{4}C$ (bottom) as a function of particle density, $n$, in the simulations introduced in \cite{soler2013}.
The values of $A_{1}$ and $C$ correspond to the definitions in Eq.~\ref{eq:A1} and Eq.~\ref{eq:C}, respectively.
The grey vertical line, drawn for reference, corresponds to $n=10^4$\,cm$^{-3}$. 
}
\label{fig:A-B}
\end{figure}
\begin{figure}[ht!]
\centerline{\includegraphics[width=0.5\textwidth,angle=0,origin=c]{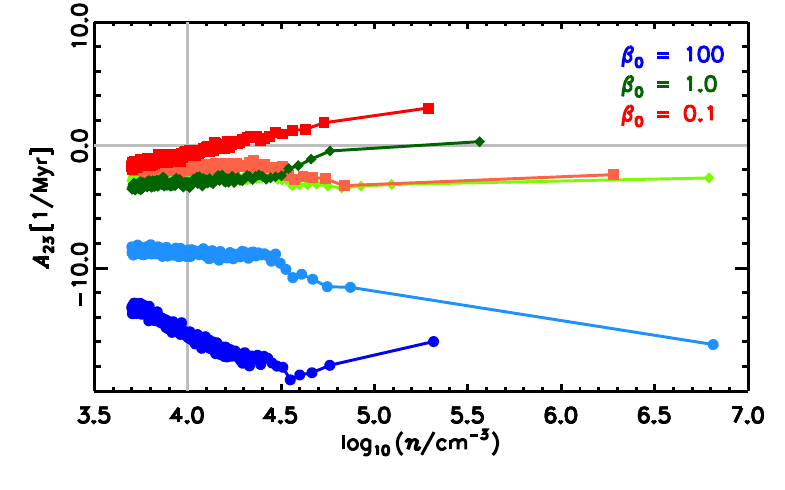}}
\vspace{-0.3cm}
\caption{Mean values of the term $A_{23}$, defined in Eq.~\ref{eq:A23diag}, as a function of particle density, $n$, in the simulations introduced in \cite{soler2013}.
The grey vertical line, drawn for reference, corresponds to $n=10^4$\,cm$^{-3}$.
}
\label{fig:A}
\end{figure}

We further quantify the source of change in relative orientation in the simulations with $\beta_{0}=1$ and $\beta_{0}=0.1$ by focusing on the behaviour of the coefficients $C$, $A_{1}$, and $A_{23}$ in Eq.~\ref{eq:costhetaCoeff}.
Fig.~\ref{fig:A-B} and Fig.~\ref{fig:A} show the values of $C$, $A_{1}$, and $A_{23}$ in bins of $n$ with equal number of voxels. 
The values shown in the figures confirm that $C$ and $A_{1}$ are considerably smaller than the values of $A_{23}$, which is expected given that the former depends on second-order spatial derivatives of the velocity field while the latter depend on first-order derivatives.
Additionally, the values of both $C$ and $A_{1}$ fluctuate around zero, thus making their mean values small with respect to $A_{23}$.

Given that $A_{23}/C \approx A_{23}/A_{1} \approx 10^{4}$, Eq.~\ref{eq:costhetaCoeff} reduces to 
\begin{equation}\label{eq:costhetaII}
\frac{d(\cos\phi)}{dt} \simeq A_{23}\cos\phi,
\end{equation}

which implies that the change in relative orientation depends mainly on the sign of $A_{23}$.
If $A_{23} < 0$, $\cos\phi$ tends to decrease towards $0$, which corresponds to $\vec{B}$ being perpendicular to $\vec{\nabla}\rho$. 
If $A_{23} > 0$, $\cos\phi$ tends to increase towards $\pm1$, which corresponds to $\vec{B}$ being parallel to $\vec{\nabla}\rho$. 


A comparison between Fig.~\ref{fig:zeta-lognh} and Fig.~\ref{fig:A} indicates that the change in relative orientation between $\vec{\nabla}\rho$ and $\vec{B}$ in the first snapshot of the $\beta_{0}=1$ and the $\beta_{0}=0.1$ simulations happens around the same $n$ values where $A_{23}$ changes its sign.
In contrast, in the $\beta_{0}=100$ simulation, where there is no change in relative orientation, the values of $A_{23}$ are always negative.
In the second snapshot, Fig.~\ref{fig:A} reveals that $A_{23}$ is negative for the equipartition and high magnetization simulations and accordingly, the perpendicular relative orientation at the highest densities is less prominent than in the first snapshot.

\subsubsection{Compression and change in relative orientation}\label{sec:PrincipalComponents}

\begin{figure}[ht!]
\centerline{
\includegraphics[width=0.48\textwidth,angle=0,origin=c]{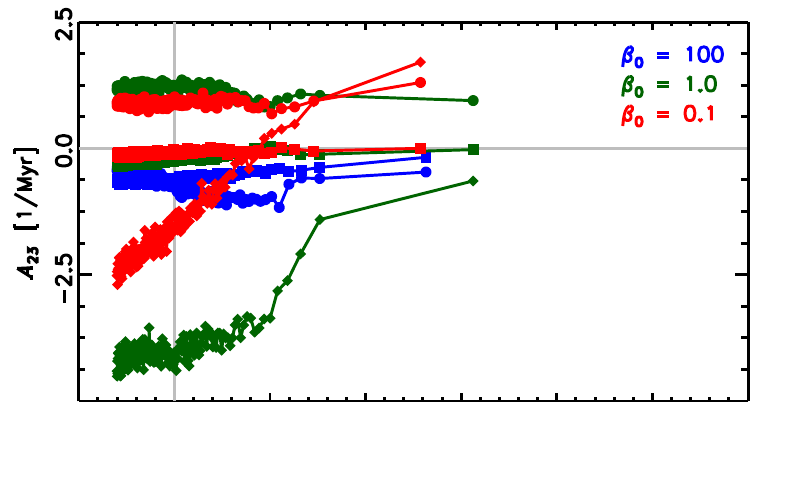}
}
\vspace{-1.2cm}
\centerline{
\includegraphics[width=0.48\textwidth,angle=0,origin=c]{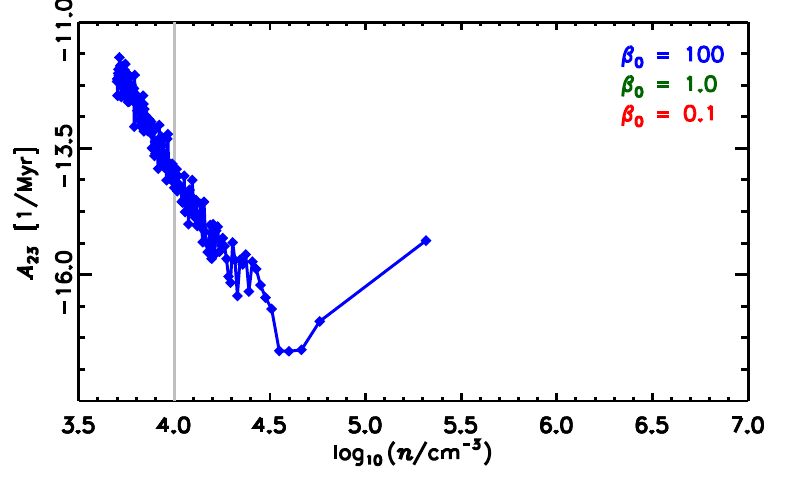}
}
\vspace{-0.3cm}
\caption{Mean values of the principal components of $A_{23}$, defined in Eq.~\ref{eq:A23tranI}, as a function of density, $n$, in the simulations introduced in \cite{soler2013}.
The component associated with the most negative eigenvalue of the symmetric part of the strain tensor, which corresponds to the dominant compressive mode, is represented by the diamonds.
These values are outside of the range in the case of the $\beta_0=100$ simulation, thus they are presented separately in the lower panel. 
The intermediate and maximum eigenvalues are represented by circles and squares, respectively.
}
\label{fig:A23eigen}
\end{figure}

Given that the tensor $(\partial_{i}v_{j}+\partial_{j}v_{i})/2$ in Eq.~\ref{eq:A23sym} is symmetric, there is a transformation $\mathcal{D}$ that diagonalises it and facilitates the interpretation of the physics included in the $A_{23}$ coefficient.
We can write Eq.~\ref{eq:A23sym} as
\begin{equation}\label{eq:A23tranI0}
A_{23}=\frac{1}{2}\mathcal{D}[\partial_{i}(v_{j})+\partial_{i}(v_{j})]\mathcal{D}^{-1}\mathcal{D}[r_{i}r_{j}-b_{i}b_{j}]\mathcal{D}^{-1},
\end{equation}
which reduces to 
\begin{equation}\label{eq:A23tranI}
A_{23}=\lambda_{i}\delta_{ij}\mathcal{D}[r_{i}r_{j}-b_{i}b_{j}]\mathcal{D}^{-1},
\end{equation}
where $\lambda_{i}$ are the three eigenvalues of $(\partial_{i}v_{j}+\partial_{j}v_{i})/2$ and $\delta_{ij}$ is the Kronecker delta. 
The eigenvector associated to the highest eigenvalue describes the axis along which the fluid parcel is mostly elongated.
The two other eigenvectors, associated to the two other eigenvalues, correspond to the directions along which the shape of the fluid parcel is either stretched or compressed, depending on their signs \citep{lai2010}.

We diagonalize the strain tensor in each position of the simulation cube and computed the $A_{23}$ terms associated to each one of its eigenvalues, as described in Eq.~\ref{eq:A23tranI}.
The mean value of each one of these terms in density bins with equal number of voxels are presented in Fig.~\ref{fig:A23eigen}.
In the low magnetization simulation, all three components of $A_{23}$ are negative and although one of them is clearly different, which can be associated to some degree to the anisotropy produced by the magnetic field, their sign implies that all of them contribute to keep the $\cos\phi=0$ configuration.
In the equipartition and high magnetization simulations, the component of $A_{23}$ associated to the most negative eigenvalue, which traces the main compressive mode, increases its mean value with increasing density.
This can be interpreted in similar terms to those described in Sec.~\ref{sec:ToyDiagonalized}: $\partial_{i}v_{i}$ is mostly negative so $A_{23}$ remains negative as long the correlation between the orientation of the magnetic field components, represented by $b_{i}b_{j}$ is smaller than that of the density gradient, represented by $r_{i}r_{j}$.
Once $b_{i}b_{j}$ becomes larger than $r_{i}r_{j}$ by the effect of the compression of the magnetized medium, the sign of $A_{23}$ changes and the system changes towards the $\cos\phi=\pm1$ configuration.

\section{Discussion}\label{sec:discussion}

\subsection{What is changing when the relative orientation between $\vec{\nabla}\rho$ and $\vec{B}$ changes?}

The clear anisotropy produced by the magnetic field in the velocity and the density distributions is expected, at least in the case of incompressible turbulence \citep{sridhar1994,goldreich1995}.
However, the reasons why compressible MHD turbulence produces structures in particular configurations with respect to the magnetic field were less clear.

\cite{hennebelle2013a} shows that non-self-gravitating filaments are a generic consequence of turbulent strain in a magnetised medium and their elongation along the magnetic field lines by effect of the Lorentz force helps to keep them coherent.
The fact that the shear modes are mostly responsible for the observed and simulated stretching of matter along the magnetic field, is confirmed by the $\cos\phi=0$ attractor solution of Eq.~\ref{eq:costheta3}, which is the dominant in the case that $\partial_{i}v_{i}$ is small.
The $\cos\phi=\pm1$ attractor solution of Eq.~\ref{eq:costheta3}, explains why the change in relative orientation is observed at the highest densities: it is produced by the compressive modes that produce the accumulations of matter.
In the light of Eq.~\ref{eq:costheta3}, the question is not what produces the relative orientation but rather what makes it change and how can we use that information to learn about the magnetic field in MCs.

\cite{chen2016} describes the transition density, from mostly $\cos\phi=0$ to $\cos\phi=\pm1$, as threshold for the gravity-driven Alfv\'{e}nic transition, from sub- to super-Alfv\'{e}nic turbulence.
Such transition in relative orientation is indeed expected, if the turbulence is sub-Alfv\'{e}nic, slow modes are important and they tend to be anti-correlated with the density field while it is the contrary for super-Alfv\'{e}nic turbulence, where fast modes are dominant.
This interpretation is not easy to generalise to the ISM, where dense structures are the result of multiple shocks induced by the super-Alfv\'{e}nic turbulence and the structure of the magnetic field may be inherited from the very large scales.
Moreover, it is not clear to what extent one can unambiguously separate regions into sub- and super-Alfv\'{e}nic turbulence.

\begin{figure}[ht!]
\centerline{\includegraphics[width=0.48\textwidth,angle=0,origin=c]{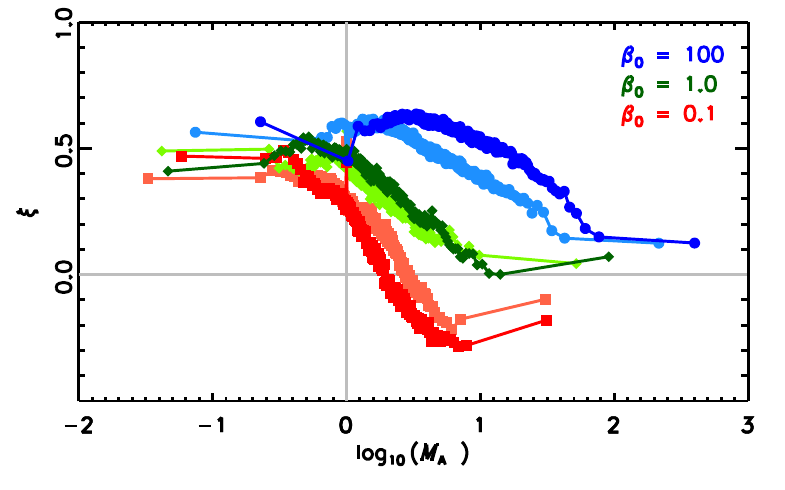}}
\vspace{-0.3cm}
\caption{Relative orientation parameter, $\xi$, as a function of the Alfv\'{e}n Mach number, $\mathcal{M}_{\rm A}$, in the simulations introduced in \cite{soler2013}.
The darker colours represent the early snapshots in the simulation and the lighter colours represent the later snapshots.
The values $\xi > 0$ correspond to $\vec{\nabla}\rho$ mostly perpendicular to $\vec{B}$ and $\xi < 0$ correspond to $\vec{\nabla}\rho$ mostly parallel to $\vec{B}$. 
The grey horizontal line is $\xi=0$, which corresponds to the case where there is no preferred relative orientation between $\vec{\nabla}\rho$ and $\vec{B}$.
The grey vertical line, drawn for reference, corresponds to $\mathcal{M}_{\rm A}=1$, which marks the border between sub-, $\mathcal{M}_{\rm A}<1$,  and super-Alfv\'{e}nic, $\mathcal{M}_{\rm A}>1$, turbulence.}
\label{fig:zeta-Mach}
\end{figure}

For the sake of the discussion, we estimated the relative orientation parameter in bins of Alfv\'{e}n Mach number, $\mathcal{M}_{\rm A}\equiv\sigma_{v}/v_{\rm A}=\mathcal{M}_{\rm S}\,\beta^{2}$, in the considered set of simulations.
If the transition between $\cos\phi=0$ to $\cos\phi=\pm1$ was related to the transition from sub-Alfv\'{e}nic to super-Alfv\'{e}nic turbulence, one should expect that the regions where $\mathcal{M}_A < 1$ would be associated to $\xi>0$ and the regions where $\mathcal{M}_A > 1$ would be associated to $\xi<0$.
Fig.~\ref{fig:zeta-Mach} shows that this is not necessarily the case in the considered simulations.
In the realisation with $\beta_{0}=100$, there is a transition from $\mathcal{M}_A < 1$ to $\mathcal{M}_A > 1$, due to the increase in velocity dispersions as a product of the gravitational collapse into denser structures, but there is no change in the sign of $\xi$.
The realisation with $\beta_{0}=0.1$, show that the $\xi < 0$ values are indeed associated with regions where $\mathcal{M}_A > 1$, but that $\mathcal{M}_A > 1$ does not necessarily imply $\xi > 0$, thus indicating that there must be another quantity that is more directly responsible for the change in relative orientation between $\vec{\nabla}\rho$ and $\vec{B}$.

Equation~\ref{eq:costheta3} and the results of the analysis presented in Sec.~\ref{sec:PrincipalComponents} suggest that the change between $\cos\phi=0$ and $\cos\phi=\pm1$, or from $\vec{\nabla}\rho$ being predominantly perpendicular to being predominantly parallel to $\vec{B}$, are related to the tensors composing the coefficient $A_{23}$, as stated by Eq.~\ref{eq:A23sym}.
If $A_{23} < 0$, the flow tends towards the $\cos\phi=0$ configuration, the change of that tendency in the case of a predominantly converging flow, $\partial_{i}v_{i} < 0$, is related to the change of sign in the tensor $(r_{i}r_{j}-b_{i}b_{j})$.
Thus, we can infer that density threshold where $\cos\phi$ changes is related to the relation between the $r_{i}r_{j}$ and the $b_{i}b_{j}$ tensors, which is clearly depent on the initial magnetization. 

\begin{figure*}[ht!]
\centerline{
\includegraphics[width=0.45\textwidth,angle=0,origin=c]{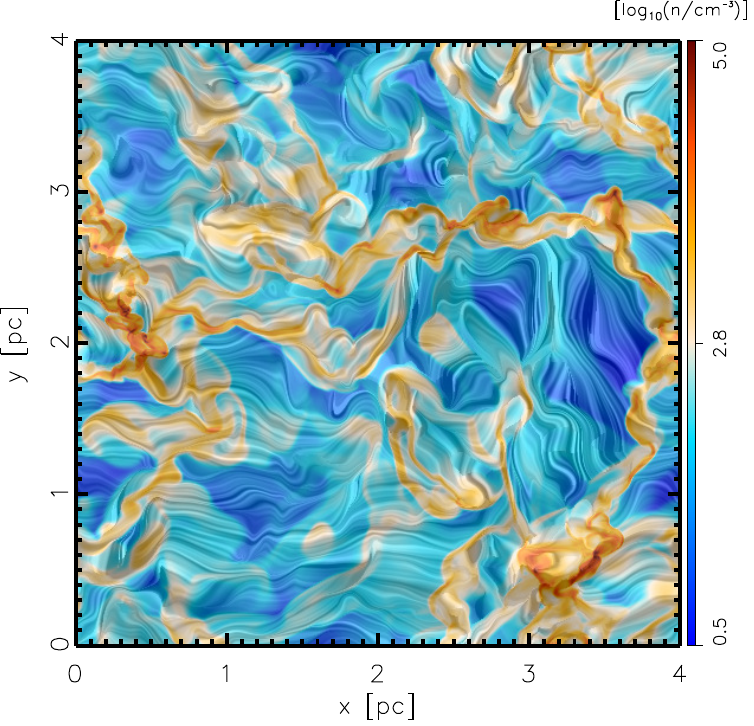}
\includegraphics[width=0.45\textwidth,angle=0,origin=c]{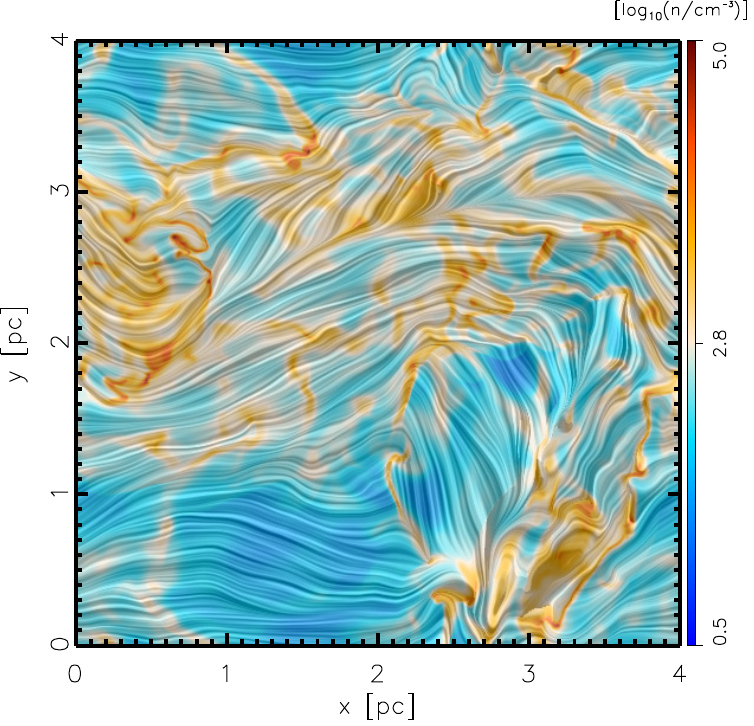}
}
\vspace{-0.3cm}
\centerline{
\includegraphics[width=0.45\textwidth,angle=0,origin=c]{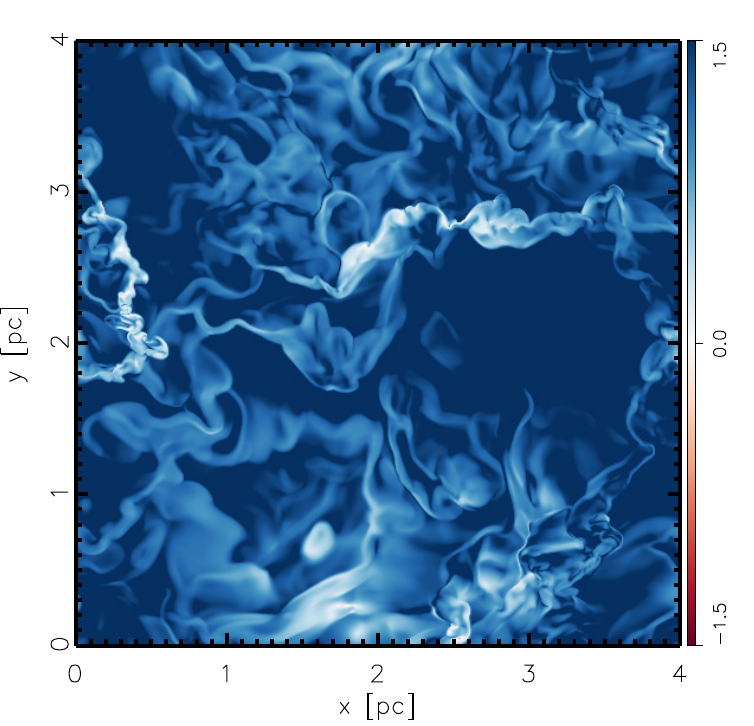}
\includegraphics[width=0.45\textwidth,angle=0,origin=c]{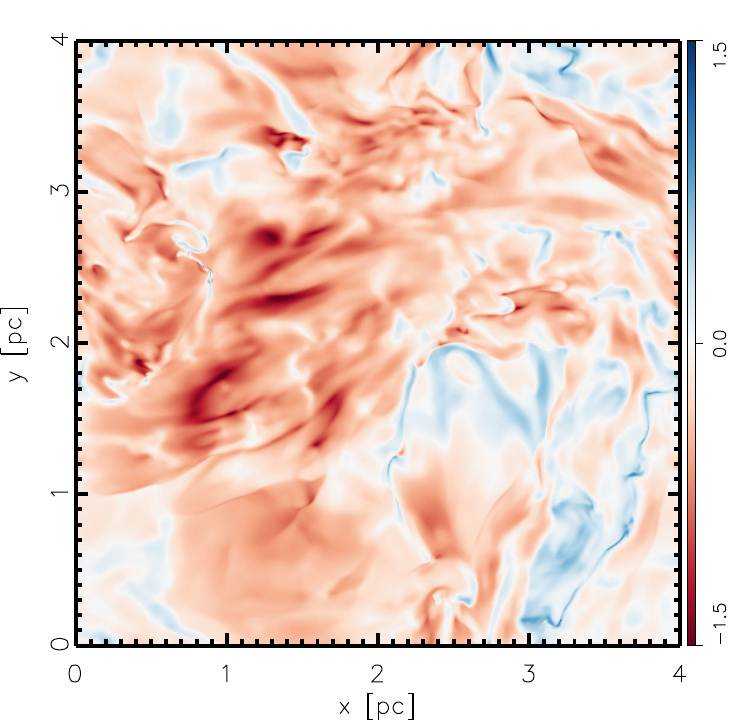}
}
\vspace{-0.3cm}
\centerline{
\includegraphics[width=0.45\textwidth,angle=0,origin=c]{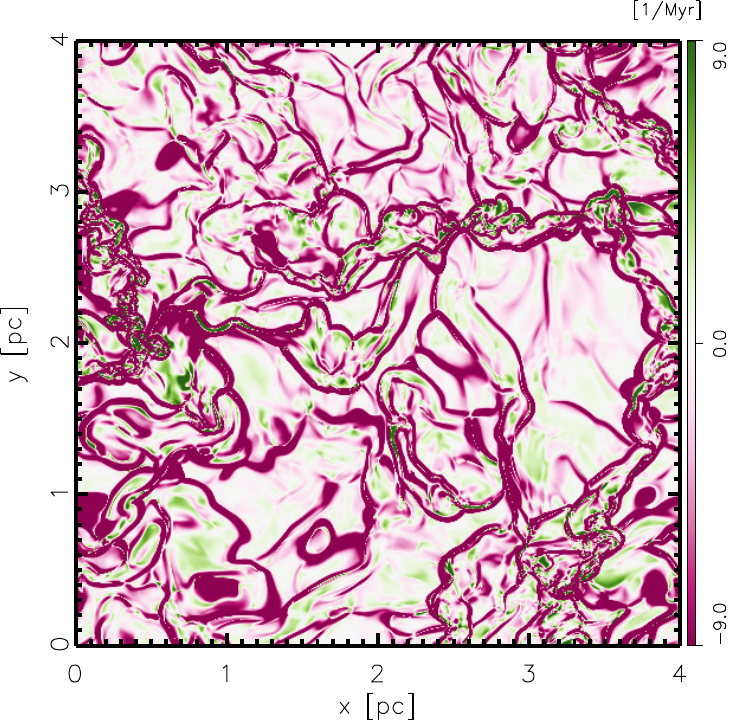}
\includegraphics[width=0.45\textwidth,angle=0,origin=c]{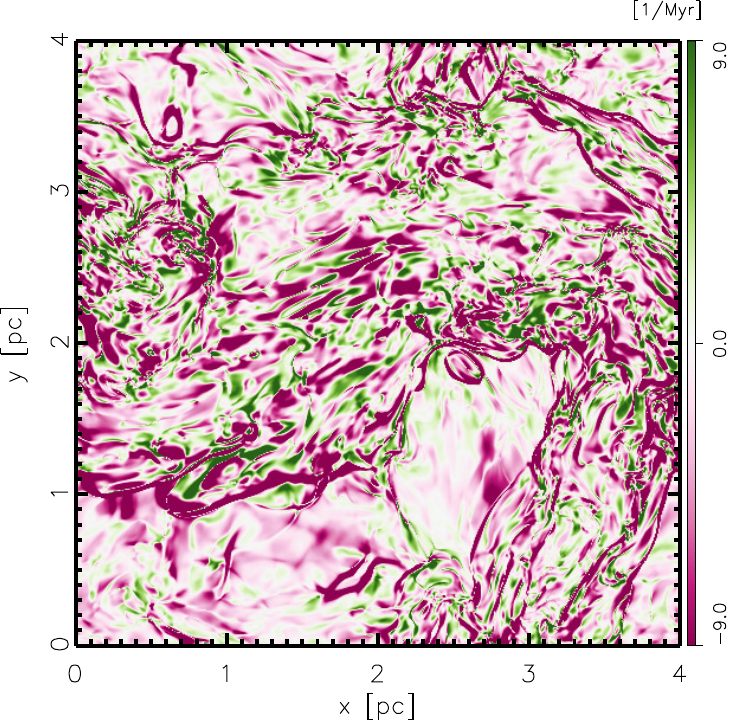}
}
\caption{Distributions of the density and magnetic field (top), the logarithm of the Alfv\'{e}n Mach number, $\mathcal{M}_{\rm A}$ (middle), and the coefficient $A_{23}$ of Eq.~\ref{eq:costhetaCoeff} (bottom) in a slice ($z=256$) of the MHD turbulence simulations used in \cite{soler2013} for initially weak ($\beta=100$, left) and strong ($\beta=0.1$, right) magnetic field in a snapshot taken at one third of the flow crossing time.
The coefficient $A_{23}$ is the term dominating the change in the relative orientation between $\vec{\nabla}\rho$ and $\vec{B}$ according to Eq.~\ref{eq:costhetaCoeff} 
}
\label{fig:Amaps256}
\end{figure*}

Fig.~\ref{fig:Amaps256} presents the distributions of $\rho$ and $\vec{B}$ orientation, $A_{23}$, and $\mathcal{M}_{\rm A}$ in a slice of the first snapshot of the simulations with weak, $\beta_0=100$, and strong $\beta_0=0.1$ initial magnetization. 
The upper panel illustrates how $\vec{B}$ clearly follows the density structure in the simulation with $\beta_0=100$ and how $\vec{B}$ is more homogeneous in the simulations where $\beta_0=0.1$.
For the sake of completeness, we also present the values of $\mathcal{M}_{\rm A}$ in the middle panel of Fig.~\ref{fig:Amaps256}  
The figure show that the values of $\mathcal{M}_{\rm A}$ in the lower density regions of the slice correspond to the initial magnetizations, that is, $\mathcal{M}_{\rm A}>1$ in the $\beta_0=100$ and $\mathcal{M}_{\rm A}<1$ in the $\beta_0=0.1$ case.
At higher densities, however, these values change and some density structures in the $\beta_0=100$ case present $\mathcal{M}_{\rm A}\approx1$ and some density structures in the $\beta_0=0.1$ case show $\mathcal{M}_{\rm A}\gtrsim1$.
Both cases, illustrate that the change in relative orientation is not straightforward to trace in some particular structures in the map, but it is rather a statistical trend that is better summarized in the analysis presented in Fig.~\ref{fig:zeta-Mach}.

Finally, the bottom panel of Fig.~\ref{fig:Amaps256} shows the distribution of $A_{23}$, the terms whose sign determines the changes in the relative orientation between $\vec{\nabla}\rho$ and $\vec{B}$.
In the case with low initial magnetization, $A_{23}$ is clearly organized in filaments that correspond to many of those seen in the density map, most likely produced as the results of shocks.
The presence of regions where $A_{23}>0$ indicates that there are zones where the relative orientation between $\vec{\nabla}\rho$ and $\vec{B}$ tends to change, although the average values in particular density ranges shows a general tendency towards the $\cos\phi=0$ configuration.
In the case with high initial magnetization, $A_{23}$ is more inhomogeneous and the density structures do not have clear counterpart in $A_{23}$.
The presence of extended regions with both positive and negative values of $A_{23}$ shows that the change in relative orientation is a dynamic process that is not localized in a few structures but rather a statistical trend that becomes evident when analyzing particular density ranges, as shown in Fig.~\ref{fig:A23eigen}.

\subsection{Relative orientation and scaling of the magnetic field with increasing density}

Observations of the Zeeman effect indicate that in the diffuse lines of sight (\nh\,$\lsim 10^{21.5}$\,cm$^{-2}$), the maximum magnetic field strength $B_{\rm max}$ sampled by HI lines does not scale with density.
In the denser regions (\nh\,$\gsim 10^{21.5}$\,cm$^{-2}$), probed by OH and CN spectral lines, the same study reports a scaling of the maximum magnetic field strength $B_{\rm max} \propto n^{0.65}$ \citep[][and references therein]{crutcher2010,crutcher2012}.
The former observation can be interpreted as the effect of diffuse clouds assembled by flows along magnetic field lines, which would increase the density but not the magnetic field strength.
The latter observation can be interpreted as the effect of isotropic contraction of weakly magnetized gas.
Probably related to these interpretations is the fact that the column density where the magnetic field strength starts scaling with increasing column density is very close to the column densities where \cite{planck2015-XXXV} identified the change in relative orientation from \bperp\ mostly parallel to the iso-\nh\ contours to mostly perpendicular.

The column density values around which the Zeeman observations show the scaling of the magnetic field with increasing density are indeed very close to the column densities where the relative orientation between the column density structures and the magnetic field, inferred from the dust polarization observations, changes from mostly parallel to mostly perpendicular.
This similarity has been identified in the colliding flows models presented in \cite{chen2016}, where is it assigned to the beginning of gravity-induced acceleration in terms of increasing gas velocity with density.
As discussed here, this may not be necessarily the case. 


\section{Conclusions}\label{sec:conclusions}

We studied the relative orientation between $\vec{\nabla}\rho$ and $\vec{B}$ using the transport equations of MHD turbulence. 
Under the assumptions of flux freezing and low magnetic diffusivity, we arrived to Eq.~\ref{eq:costheta3}, which is an expression that describes the evolution of $\cos\phi$, the cosine of the angle between $\vec{\nabla}\rho$ and $\vec{B}$.

From the study of Eq.~\ref{eq:costheta3} we conclude:
\begin{enumerate}
\item The configuration where $\cos\phi=\pm1$ is a generic attractor, a configuration towards which the system tends to evolve.
\item The configuration where the $\cos\phi=0$ constitutes another attractor. 
\item The changes in the relative orientation are produced by the coupling of divergence of the velocity field, $\partial_{i}v_{i}$, and the $r_{i}r_{j}$ and $b_{i}b_{j}$ tensors, defined in Eq.~\ref{eq:defri} and Eq.~\ref{eq:defbi}, which correspond to the correlations of the density gradient and magnetic field orientations, respectively.
\end{enumerate}

Using the simulations of MHD turbulence used in \cite{soler2013}, we show that:
%
\begin{enumerate}
\item The configuration $\cos\phi=0$ is dominant at all densities in the quasi-hydrodynamic simulation ($\beta_{0}=100$) and in all but the highest densities in the initially equipartition ($\beta_{0}=1.0$) and high-magnetization ($\beta_{0}=0.1$) simulations, as illustrated in Fig.~\ref{fig:zeta-lognh}
\item The $\cos\phi=\pm1$ is only present in the regions where $\partial_{i}v_{i} < 0$ in the $\beta_{0}=1.0$ and $\beta_{0}=0.1$ simulations, as show in Fig.~\ref{fig:zeta-divV}.
\item The density over which $\cos\phi$ changes from $0$ to $\pm1$ in the $\beta_{0}=1.0$ and $\beta_{0}=0.1$ simulations corresponds to that where the term $A_{23}$ changes its sign, as illustrated in Fig.~\ref{fig:A}
\item In the case of $\partial_{i}v_{i} < 0$, the change of sign in $A_{23}$ occurs when the $b_{i}b_{j}$ tensor dominates over $r_{i}r_{j}$.
\item The density over which $A_{23}$ changes sign, bringing the system from the $\cos\phi=0$ to the $\cos\phi=\pm1$, depends on the strength of the initial magnetic field.
\item The changes in $\cos\phi$ are mainly related to the compressive modes of the strain tensor and their coupling to a relatively strong magnetic field, as illustrated in Fig.~\ref{fig:A23eigen}, and not to a clear transition in the Alfv\'{e}n Mach number, as shown in Fig.~\ref{fig:zeta-Mach}.
\end{enumerate}

These results indicate that, once the projection effects are properly accounted for, the observations of the relative orientation between column density structures and the projected magnetic field can be interpreted as follows: 
\begin{enumerate}
\item The observed low-density structures aligned with the magnetic field are spontaneously produced in regions where shear motions are dominant over compression.
\item The observed change in relative orientation between the \nh\ structures and \bperp\ towards molecular clouds is an indication of compressive motions, that is, $\partial_{i}v_{i}< 0$, which can be the result of either gravitational collapse or converging flows.
\item The density threshold at which the relative orientation between the \nh\ structures and \bperp\ changes from mostly parallel to mostly perpendicular depends on the field strength and not necessarily on the strength of the compressive motions.
\end{enumerate}

We have shown that the relative orientation between density structures and the magnetic field are related to the dynamics of MHD turbulence, particularly to the contraction motions and the degree of magnetization.
The clear astrophysical implication of our calculations is that the current polarization observations suggest that the magnetic field must play an important role in the assembly of the parcels of gas that produce molecular clouds.
While the direct estimation of the magnetic field strength from observations of the relative orientation between \nh\ structures and \bperp\ remains elusive, it is clear that the study of this and other statistical correlations is crucial for constraining the range of scales and densities where the magnetic field is shaping the structure of the ISM.

\begin{acknowledgements}
This work was possible through the funding from the European Research Council under the European Community's Seventh Framework Programme (FP7/2007-2013 Grant Agreement no. 306483 and no. 291294).
JDS acknowledges the support from the European Research Council under the Horizon 2020 Framework Program via the ERC Consolidator Grant CSF-648505.
We thank the following people who helped with their encouragement and conversation: Francois Boulanger, Ralf Klessen, Henrik Beuther, and Jouni Kainulainen.
\end{acknowledgements}

\bibliographystyle{aa}
\bibliography{RelativeOrientation.bbl}
\clearpage

\appendix

\section{Relative orientation between $\vec{B}$ and $\vec{v}$}\label{app:BandV}

Following the a similar procedure to that described in Sec.~\ref{sec:MHDeq}, we obtained and expression for the relative orientation between the magnetic field, $\vec{B}$, and the velocity, $\vec{v}$.  
We begin by considering the Cauchy momentum equation in a magnetized fluid, 
\begin{equation}\label{eq:momentum}
\frac{dv_{i}}{dt} = -\frac{\partial_{i}P}{\rho} + \frac{B_{j}}{\rho}\partial_{j}B_{i}-\frac{B_{j}}{\rho}\partial_{i}B_{j}\,.   
\end{equation}
Then by combining the continuity and the Faraday equation, we obtain
\begin{equation}\label{eq:continuityAndFaraday}
\frac{d (B_{i}/ \rho)}{dt} = \frac{B_{j}}{\rho}\partial_{j}v_{i}.  
\end{equation}

We define
\begin{equation}\label{eq:BV}
\cos \theta = \frac{v_{i}B_{i}}{(v_{k}v_{k})^{1/2}(B_{k}B_{k})^{1/2}} 
\end{equation}
and compute
\begin{equation}\label{eq:BV1}
\begin{split}
\frac{d\cos^{2}\theta}{dt} &= \frac{2}{(v_{k}v_{k})^{2} (B_{l} B_{l})^{2}}\Bigg[  {d (v_i B_i) \over dt} (v_{j} B_{j}) (v_{m}v_{m})(B_{n}B_{n})\\
& - \frac{dB_{n}}{dt}B_{n}(v_{i}B_{i})^{2}(v_{m}v_{m}) - \frac{dv_{m}}{dt}v_{m}(B_{n}B_{n}) (v_{i}B_{i})^2 \Bigg], 
\end{split}	
\end{equation}
which leads to 
%
\begin{equation}\label{eq:BV2}
\begin{split}
\frac{d\cos^2 \theta}{dt}=& \frac{2}{(v_{k}v_{k})^{2} (B_{l}B_{l}/ \rho^2)^{2} }\Bigg[\\  
& \Big( - \partial_{i} P \frac{B_{i}}{\rho^2} + \frac{B_{p}}{\rho}\partial_{p}v_{i}v_{i}  \Big) \Big( {v_{j}B_{j} \over \rho} \Big) (v_{m}v_{m}) \Big( {B_{n}B_{n} \over \rho^2} \Big) \\
& - \Big( {B_{n} \over \rho} \partial _{n} v_{i}\Big) \frac{B_{i}}{\rho}  \Big(v_{k} {B_{k} \over \rho} \Big)^2  (v_{m}v_{m}) \\
& - \Big(\frac{B_{j}}{\rho}\partial_{j}B_{i}v_{i}  - \frac{\partial_{i} P}{\rho} v_{i}  -  {B_j \over \rho}  \partial_i B_j v_i \Big) \left( {B_{n}B_{n} \over \rho^2} \right) \left( \frac{v_{m}B_{m}}{\rho} \right)^2 \Bigg].
\end{split}	
\end{equation}

Finally, using the definitions, 
\begin{equation}
\begin{split}
u_i &\equiv \frac{v_{i}}{(v_{k} v_{k})^{1/2}}, \\
b_i &\equiv {B_i \over (B_k  B_k)^{1/2} },
\end{split}	
\end{equation}
we obtain
%
\begin{equation}
\begin{split}
\frac{d\cos\theta}{dt} =& - \frac{\partial _{i}P}{\rho (v_{k} v_{k})^{1/2}}(b_{i}-u_{i}\cos\theta) \\
&+b_{j}\frac{\partial_{j}v_{i}}{(v_{k}v_{k})^{1/2} }(u_{i} - b_{i}\cos\theta) \\
&-\frac{b_{j}\partial_{j}B_{i} - b_{j}\partial_{i}B_{j}}{ (B_k B_k)^{1/2} }u_{i}\cos\theta,
\end{split}	
\end{equation}
%
which can be written as
\begin{equation}\label{eq:Bv}
\begin{split}
\frac{d\cos\theta}{dt} =& \left[ -\frac{\partial_{i}P}{\rho v} +  \left(\frac{ B_{j}\partial_{j}B_{i} - B_{j}\partial_{i}B_{j}}{\rho v}\right) + \partial_{i}v \right]\left( b_i - u_i \cos \theta \right) \\
&+  \frac{1}{2}(\partial_{j}v_{i} + \partial_{i}v_{j})[v_{i}v_{j} - b_{i}b_{j}]\cos\theta,  
\end{split}	
\end{equation}
where $v\equiv(v_{k}v_{k})^{1/2}$.

If $\cos\theta=\pm 1$, that is to say $b_i=\pm v_i$, Eq.~\ref{eq:Bv} implies that $d\cos\theta /dt =0$, as its third term vanishes because it corresponds to the Lorenz force times the velocity, which is proportional to \vec{B}.
This implies that alignment between $\vec{B}$ and $\vec{v}$ is expected since $\cos\theta=\pm1$ constitutes an attractor, like in the case of the relative orientation between $\vec{\nabla}\rho$ and $\vec{B}$. 
However, unlike Eq.~\ref{eq:costheta3}, the configuration $\cos\theta=0$ is not an attractor due to the first term on the right-hand side of Eq.~\ref{eq:Bv}, which account for the forces. 

\begin{figure}[ht!]
\vspace{-0.1cm}
\centerline{\includegraphics[width=0.48\textwidth,angle=0,origin=c]{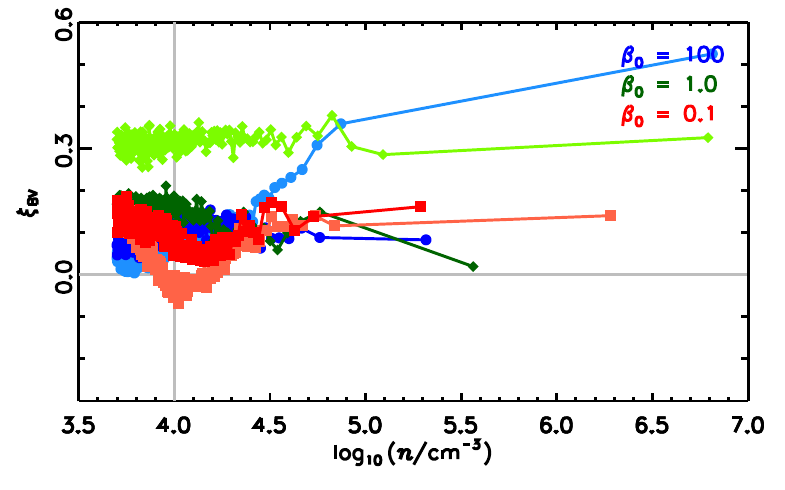}}
\vspace{-0.5cm}
\caption{Relative orientation parameter, $\xi_{\rm BV}$, as a function of particle density, $n\equiv\rho/\mu$, in the simulations used in \cite{soler2013}.
The values of $\xi_{\rm BV}$ correspond to the relative orientation between $\vec{B}$ and $\vec{v}$ in $n$-bins with equal number of voxels, all with $n>500$\,cm$^{-3}$.
The values $\xi_{\rm BV} > 0$ correspond to $\vec{B}$ mostly perpendicular to $\vec{v}$ and $\xi_{\rm BV} < 0$ correspond to $\vec{B}$ mostly parallel to $\vec{v}$.
The grey horizontal line is $\xi_{\rm BV}=0$, which corresponds to the case where there is no preferred relative orientation between $\vec{B}$ and $\vec{v}$.
The colours and the symbols represent the initial magnetization values quantified by $\beta_{0}$.
The darker colours represent the early snapshots in the simulation and the lighter colours represent the later snapshots.
The grey vertical line, drawn for reference, corresponds to $n=10^4$\,cm$^{-3}$.
}
\label{fig:zetaBV-lognh}
\end{figure}

The predominant alignment between $\vec{B}$ and $\vec{v}$ has been previously reported in \cite{boldyrev2006}, \cite{matthaeus2008}, and \cite{banerjee2009}, and more recently studied in the context of MC formation in \cite{iffrig2017}.
However, Eq.~\ref{eq:Bv} constitutes a novel general expression that link the relative orientation between the two vectors.
It relates the relative orientation to the underlying physical conditions, which is potentially useful in the interpretation of observations such as those of the magnetically aligned velocity anisotropy reported in \cite{heyer2008} and the correlations between the \bperp\ and the line-of-sight velocity gradients reported in \cite{yuen2017}.
Additionally, it provides reference to the assumptions behind methods for the estimation of the magnetic field strength, such as those presented in \cite{houde2009} and \cite{gonzalezcasanova2017}.

For the sake of illustration, we present the analysis of the relative orientation between $\vec{B}$ and $\vec{v}$ in the \cite{soler2013} simulations in Fig.~\ref{fig:zetaBV-lognh}.
There we show that in the range $n>500$\,cm$^{-3}$, the alignment between $\vec{B}$ and $\vec{v}$ is prevalent under the physical conditions included in those simulations, although the trends for increasing density are not as uniform as in the case of $\vec{\nabla}\rho$ and $\vec{B}$.
These trends are different in the first and the second snapshot as a consequence of the decay of the turbulence: stronger shocks tend to shift the alignment of $\vec{B}$ and $\vec{v}$, as
the transverse component of the magnetic field is amplified and the velocity tends to be perpendicular to it \citep{passot1995}. 


\raggedright

\end{document}